\documentclass[%
reprint,
 amsmath,amssymb,
 aps,
]{revtex4-2}

\usepackage{graphicx}% Include figure files
\usepackage{dcolumn}% Align table columns on decimal point
\usepackage{bm}% bold math
\usepackage{color}
\usepackage{subcaption}
\usepackage{xcolor}
\usepackage{soul}
\usepackage{braket}
\usepackage{graphicx}
\usepackage{caption}
\usepackage{ragged2e}

\begin{document}
\preprint{APS/123-QED}

\title{Xenon-metal pair formation in UO$_2$ investigated using DFT+$U$}% Force line breaks with \\
\author{Linu Malakkal$^1$}
 \altaffiliation[ ]{linu.malakkal@inl.gov}%Lines break automatically or can be forced with \\
\author{Shuxiang Zhou$^1$}%
\author{Himani Mishra$^1$}%
\author{Mukesh Bachhav$^1$}%
\author{Jia Hong Ke$^1$}%
\author{Chao Jiang$^1$}%
\author{Lingfeng He$^2$}%
\author{Sudipta Biswas$^1$}%=

\affiliation{$^1$Computational Mechanics and Materials Department, Idaho National Laboratory, Idaho Falls, ID 83415, USA}
\affiliation{$^1,2$Department of Nuclear Engineering, North Carolina State University, Raleigh, NC, 27695, USA}

\date{\today}% It is always \today, today,

\begin{abstract}
 
A recent experimental study on a spent uranium dioxide (UO$_2$) fuel sample from Belgium Reactor 3 (BR3) identified a unique pair structure formed by the noble metal phase (NMP) and fission gas (xenon [Xe]) precipitate. However, the fundamental mechanism behind this structure remains unclear. The present study aims to provide an understanding of the interaction between five different metal precipitates (molybdenum [Mo], ruthenium [Ru], palladium [Pd], technetium [Tc], and rhodium [Rh]) and the Xe fission gas atoms in UO$_2$, by using density functional theory (DFT) in combination with the Hubbard $U$ correction to compute the formation energies involved. All DFT+$U$ calculations were performed with occupation matrix control to ensure antiferromagnetic ordering of UO$_2$. The calculated formation and binding energies of the Xe and solid fission products in the NMP reveal that these metal precipitates form stable pair structures with Xe. Notably, the formation energy of Xe-metal pairs is lower than that of the isolated single defects in all instances, with Pd and Mo showing the most favourable binding energy, likely accounting for the observed pair structure formation.

%\begin{description}
%\end{description}
\end{abstract}

%Use showkeys class option if keyword
\maketitle

%\tableofcontents

\section{\label{sec:level1}Introduction \protect}

Irradiation of nuclear fuels produces various fission products, which tend to segregate within different regions of the fuel's microstructure, based on their diffusivity and reactivity. Among the various fission products, a well-documented segregated phases in uranium dioxide (UO$_2$) \cite{BRADBURY1965} and mixed oxide fuels \cite{OBOYLE_1970} are metallic precipitates primarily composed of molybdenum (Mo) and ruthenium (Ru), with smaller amounts of technetium (Tc), rhodium (Rh), and palladium (Pd). The literature refers to these metallic precipitates by various names, including white inclusion \cite{BRAMMAN1968,BRADBURY1965}, epsilon particles \cite{IMOTO_1986,KAYE_2007}, fission product alloys, the five metal precipitates (5MPs) (i.e., Mo, Ru, Tc, Rh, Pd) \cite{SCHEELE2012}, and the noble metal phase (NMP) \cite{KLEYKAMP_1985_2}. Recent studies have brought to light new insights: Buck et al.~\cite{BUCK_2015} reported uncertainties regarding the degree of crystallinity of these precipitates, whereas Kessler et al.~\cite{KESSLER2020} provided evidence for a separate non-metallic phase associated with the metallic particles. Additionally, Pellegrini et al. \cite{Pellegrini_2019} and Kessler et al.~\cite{KESSLER2020} highlighted the presence of tellurium alongside the 5MPs. Given these findings, the term ``NMP'' is considered the most appropriate nomenclature, and we adopt this terminology so as to maintain continuity with recent scholarly work. 
NMP typically segregate at the grain boundaries of UO$_2$ \cite{BUCK_2015}, potentially causing fuel pellet swelling or alterations in the brittleness of surrounding materials \cite{RAY_1992}, thereby impacting the mechanical properties and performance of nuclear fuel. Further, Cui et al.~\cite{CUI_2010} demonstrated that these particles extracted from spent fuel components, when suspended in an aqueous solution, can catalytically contribute to the reduction of actinides, promoting immobilization of the waste form, as is critical for effective nuclear waste management. Thus, the significance of NMP has led numerous researchers to conduct both experimental and theoretical studies to understand their formation mechanisms, size distributions, and behaviors. 

The theoretical studies have primarily utilized CALPHAD \cite{saunders1998calphad} and \emph{ab initio} techniques to comprehend the thermodynamics of mixing in metal alloys under varying oxygen (O) potentials. Initially, researchers reduced quinary systems to binary \cite{CHATTOPADHYAY_1986,KLEYKAMP_1985_2,KLEYKAMP_1989,KELM_1990,MATSUI_1989, MATSUI_1990, KISSAVOS_2005,Kissavos_2007} or ternary \cite{DWIGHT_1985,CORNISH_1997} subsets in order to elucidate the thermodynamic stability of the NMP. Building on these studies, Kaye et al.~\cite{KAYE_2007} developed a comprehensive thermodynamic model for the quinary system. Middleburgh et al.~\cite{Middleburgh_2015} were the first to use density functional theory (DFT) to model the hexagonal structure of NMP, reporting vacancy formation energies and investigating the behavior of extrinsic defects (e.g., xenon [Xe] and iodine) in the hexagonal phase. Furthermore, King et al.~\cite{KING2017} explored the stability, partitioning, and partial ordering of NMP systems as a function of temperature and composition. Kleykamp et al.~\cite{KLEYKAMP_1985_jnm, KLEYKAMP_1985_jnm2} investigated the composition and structure of fission product precipitates, noting that Mo is the most readily oxidized constituent of NMPs and reacts with any excess oxygen in the system. As burnup increases, oxygen is released in the oxide fuel, leading to a reduction in Mo concentration, thus making Mo content a reliable measure of burnup.

Recently, we performed advanced characterization of a spent UO$_2$ fuel sample with an average burnup was approximately 40 MWdkg$^{-1}$, from Belgium Reactor 3 (BR3) \cite{Lingfeng_Ditter_2022,Lingfeng_YUAN_2021}. Our investigation revealed that the fission product Xe forms intriguing pair structures with NMP. However, the fundamental mechanism behind these intriguing pair structures remains unknown. Although several studies have explored the stability of the NMP, the interactions between NMP and fission gases remain poorly understood. To address this knowledge gap, we employed first-principles calculations to investigate possible defect pair formations in UO$_2$. Theoretical studies have previously utilized DFT simulations to understand defect energetics and diffusion of various fission products in actinide dioxides \cite{YUN_2008,Nerikar_2009,Crocombette_2011,Andersson_PRB_2011,Thompson_2011_PRB,Minki_2012, COOPER_2018,REST_2019_JNM,SINGH_2024_JNM,Neilson_2024}. For example, Yun et al.\cite{YUN_2008} investigated the atomic diffusion mechanism of Xe in UO$_2$ through vacancy-assisted diffusion, calculating the incorporation, binding, and migration energies using DFT with spin-orbit coupling (SOC). They found that an oxygen vacancy lowers the migration energy of a uranium vacancy and suggested that the strain energy of Xe significantly contributes to the clustering of vacancies, driving the vacancy-assisted diffusion of Xe in UO$_2$. Nerikar et al.\cite{Nerikar_2009} used DFT+$U$ calculations to study the stability of neutral and charged intrinsic point defects, finding that predicted equilibrium properties and defect formation energies for neutral defect complexes align well with experimental trends, though values for charged complexes were lower than measured. Thompson et al.~\cite{Thompson_2011_PRB} conducted DFT+$U$ calculations to explore the energetics of various defects in UO$_2$, including noble gases (He, Ne, Ar, Kr, Xe), Schottky defects, and the interaction between these defects. Andersson et al.~\cite{Andersson_PRB_2011} proposed mechanisms governing fission gas evolution by analyzing Xe solution thermodynamics, migration barriers, and the interaction of dissolved Xe atoms with uranium, demonstrating that Xe diffusion predominantly occurs via a vacancy-mediated mechanism. Hong et al.~\cite{Minki_2012} used DFT to investigate the solubility and clustering of Ru fission products in UO$_2$ suggesting that metallic dimers in Schottky defects likely form the nucleus of metallic precipitates. Cooper et al. \cite{COOPER_2018} integrated DFT calculations of defect energies with empirical potential calculations of defect vibrational entropy to analyze the point defect concentrations as functions of oxygen partial pressure, temperature, and composition in UO$_{2 \pm x}$. Shilpa et al.~\cite{SINGH_2024_JNM} recently examined the formation and migration of various neutral defects such as vacancies, interstitials, antisites, Schotkky defects, and Frenkel defects in actinide oxides. Building on the work by Hong et al.~\cite{Minki_2012} on the solubility and clustering of Ru fission products in UO$_2$, we utilized the DFT+$U$ approach to understand the formation of various metal/fission-gas pair structures. We specifically characterized the formation energies, binding energy, and charge transitions of each Xe-metal pair structure, for metals of Tc, Mo, Pd, Ru, and Rh. Although it is well-established that charged defect play a crucial role in predicting specific experimental observations of UO$_2$ under certain irradiation conditions \cite{Crocombette_2011, Nerikar_2009, COOPER_2018, Neilson_2024}, this study focuses exclusively on neutral defects, because the focus of this work centers on formation energies and binding energies, and it is recognized that charged defects generally lowers the predicted energy values compared to neutral defects \cite{Nerikar_2009}. Consequently, while restricting our analysis to neutral defects may result in slight variations in the energy values, but will not comprise the overall conclusion of this study. For the 5MPs, our DFT+$U$ calculations identified the elements that are most energetically favorable to forming these pair structures. We anticipate that our results will inform future discussions regarding the impact of defect formation on the performance and stability of UO$_2$. Furthermore, we expect that this work will stimulate future studies under various conditions encountered in nuclear reactors, ultimately advancing our understanding of defect dynamics in nuclear materials.

%%%%Experimental details
\section{Experimental Details}
A focused ion beam section was extracted from the rim area of spent BR3 light-water reactor fuel with an average burnup of 4.5 at$\%$, equivalent to approximately 40 MWd/kgU \cite{Steven_HERRMANN_2007}. The cross-sectional sample for transmission electron microscopy was prepared using a Quanta 3D Field Emission Gun (FEG) focused ion beam system. Atomic-resolution microstructural characterization was conducted using a Titan Themis 200 transmission electron microscope (TEM) in scanning TEM (STEM) mode at Idaho National Laboratory. The spent UO$_2$ fuel sample was characterized via TEM at the Hot Cell Examination Facility and the Irradiated Materials Characterization Laboratory at Idaho National Laboratory \cite{Sadman_2024}. 
\begin{figure}[h!]
    \centering
    \includegraphics[width=0.95\columnwidth]{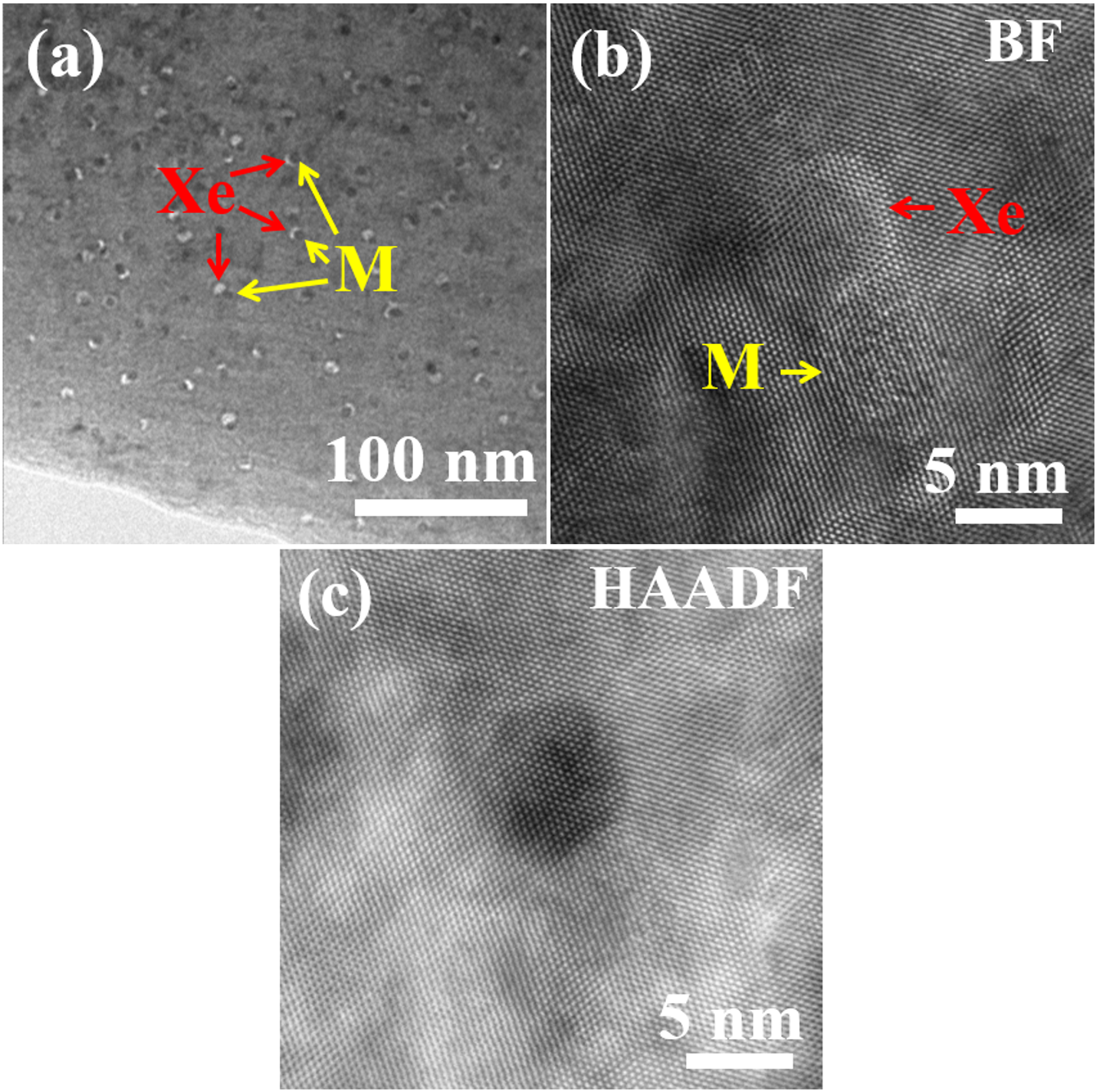}
    \caption{(a) Bright field transmission electron microscopy (TEM) image, atomic resolution (b) Bright-field (BF) image and (c) High-angle angular dark-field (HAADF) STEM images showing the pair structure of fission gas bubbles Xe and NMP (M).}
    \label{Fig1.png}
\end{figure}

Fig.~\ref{Fig1.png}(a) shows the low mag TEM image of pair structure between fission Xe/Kr gas bubbles (bright features) and NMP (M, dark features). Figs.~\ref{Fig1.png}(b) and (c) clearly illustrates the atomic-scale structure of a pair, with both Xe/Kr and NMP having a size of approximately 7 nm. The bright contrast observed in the bright-field image and the dark contrast in the high-angle annular dark-field image (i.e., Z-contrast) indicate a lower atomic density in the fission gas bubble as compared to the surrounding UO$_2$ matrix. The embedded NMP disrupts the UO$_2$ matrix, leading to lattice strain, which affects the contrast of the U atom columns (Figs.~\ref{Fig1.png}(b)). Though the crystal structure of the NMP is not fully resolved, it does not form a fully coherent interface with the UO$_2$ matrix.

Atom Probe Tomography (APT) analysis was conducted on UO$_2$ fuel to elucidate microstructural changes and the distribution of fission products. Detailed information on the APT instrument, sample preparation, and 3D reconstruction methodology is provided in the supplementary information (SI). The APT analysis site was strategically positioned adjacent to the TEM site to ensure that the same region of interest was utilized for both microstructural and chemical analysis.

\begin{figure}[h!]
    \centering
    \includegraphics[width=0.95\columnwidth]{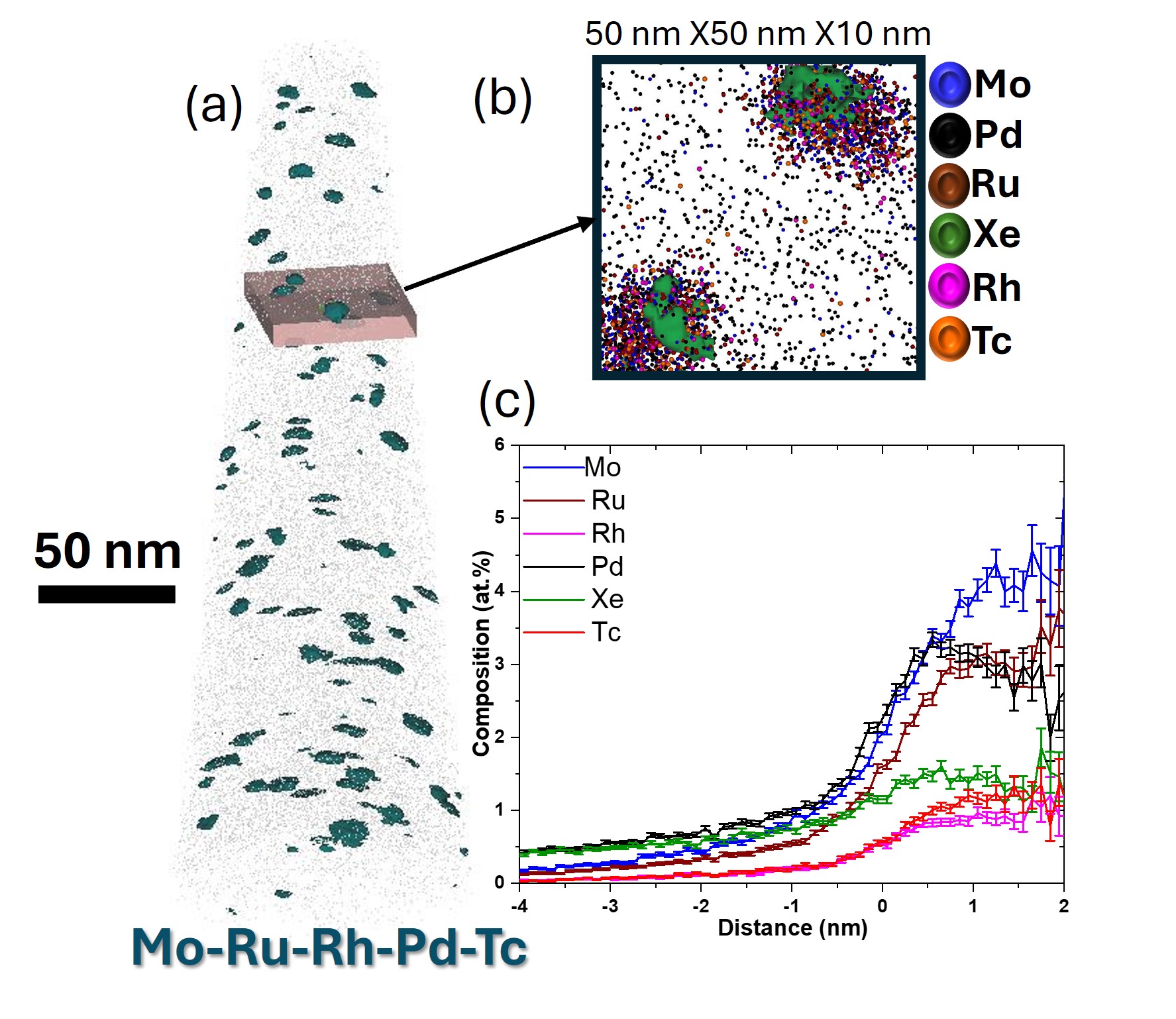}
   \caption{(a) Representative 3D reconstruction of the  UO$_2$ tip, (b) distribution of Mo-Tc-Ru-Rh-Pd and an iso-concentration contour of Xe in a sub-volume, (c) proxigram from precipitates taken from figure (a)}
    \label{Fig2}
\end{figure}

Fig.~\ref{Fig2}(a) presents a representative 3D reconstruction image of a tip, featuring an iso-concentration map that illustrates Mo-Tc-Ru-Rh-Pd precipitates uniformly distributed throughout the analyzed volume, with sizes ranging between 2-3 nm in radius. Fig.~\ref{Fig2}(b) depicts a sub-volume of a 50 nm x 50 nm x 10 nm cube, highlighting the distribution of Mo-Tc-Ru-Rh-Pd and an iso-concentration contour of Xe. Fig.~\ref{Fig2}(c) displays a proximity histogram (proxigram) plot for all precipitates, facilitating the analysis of the chemistry across all precipitates shown in Figure 2(a). It is important to note that local magnification effects in APT may impact the quantitative analysis of the size, shape, and chemistry of the five metal precipitates and Xe, due to differences in the field evaporation behavior of the elements in the UO$_2$ matrix \cite{APT_1_vurpillot, APT_2_Marquis, APT_3_Geuser}. Nonetheless, the concentrations of Mo, Pd, and Ru are the most significant. The APT analysis reveals distinct segregation of the five metal precipitates along with Xe, corroborating the findings from TEM analysis.
\section{Simulation details}
In this work, DFT calculations were carried out using the projector augmented-wave method \cite{blochl_projector_1994, kresse_ultrasoft_1999}, as implemented in the Vienna ab initio Simulation Package (VASP) code \cite{kresse_ab_1993, kresse_efficient_1996}. The exchange correlation functional used was the generalized gradient approximation formulated by Perdew, Burke, and Ernzerhof \cite{perdew_generalized_1996}. To approximate the strong correlation of 5$f$ electrons in UO$_2$, the rotationally invariant DFT+$U$ approach \cite{dudarev_electron-energy-loss_1998} with $U=4$ eV was employed. Spin-orbit coupling was also included in all calculations. While the ground state of UO$_2$ at $0$ K is in a noncollinear structure of 3\textbf{k} antiferromagnetic (AFM) state \cite{burlet_neutron_1986,ikushima_first-order_2001,blackburn_spherical_2005,wilkins_direct_2006}, it is challenging to converge all the defected structures into the 3\textbf{k} AFM state, so the 1\textbf{k} AFM state was applied in this work as an approximation \cite{Andersson_PRB_2011, Thompson_2011_PRB, crocombette2012influence, dorado2013advances}. To converge all calculations to the designed 1\textbf{k} AFM state, the occupation matrix control technique was applied. The VASP code was customized to monitor and initialize the occupation matrices during the calculation \cite{zhouCapturingGroundState2022a}, and the initial values of the occupation matrices were taken from Ref.~\cite{zhouCapturingGroundState2022a} (i.e., $\mathbb{S}_0$ of the 1\textbf{k} AFM state).

The formation energy of the Xe and five metal elements in the NMP was modeled using a 2$\times$2$\times$2 supercell with 96 atoms for the geometrical optimizations, and the cell volume was kept constant. Although the supercell size is admittedly small,
increasing the size in DFT for UO$_2$ is expensive due to utilization of DFT+$U$ and spin-orbit coupling in a noncollinear structure. It is also worth noting that the same supercell sizes were used by several researchers to tackle similar problems \cite{Minki_2012}. We implemented a  2$\times$2$\times$2 Monkhorst–Pack k-point mesh \cite{Kpoint_1976}, which is sufficient to avoid significant numerical error \cite{Minki_2012}. For all calculations reported herein, the cutoff energy for the plane waves was 550 eV. The convergence criteria for the energy difference was 10$^{-6}$ eV/atom, and for the residual forces less than 10$^{-3}$ eV/{\AA}. 

In this study, we present the formation energy as a crucial property for quantifying the ease with which the Xe-M precipitates form---specifically considering the neutral defects at substitutional uranium sites. The defect formation energy (E$_f$(X)) for a neutral defect is calculated as
\cite{Defect_pair_2022}:
\begin{equation}\label{eq1}
    E_f(X) = E(X) - E_0 - \sum_i{n_i^X(E_i+\mu_i)}
\end{equation}
where $E(X)$ is the total energy of the defected system, $E_0$ is the total energy of the pristine system, $n_i^X$ is the number of atoms of species $i$ that were changed according to the defect $X$ (positive if atoms are added, negative if atoms are removed), $E_i$ is the energy per atom in the elemental phase, and $\mu_i$ is the chemical potential of species $i$.
To determine the energy per atom in the elemental phase, we calculated the total energy of the atom in its standard state as reference system, instead of treating it as an isolated atom. In this case, the total energy of a single atom is defined as the energy per atom in each reference system. For instance, to calculate the energy per uranium atom, we used the reference state of uranium in bulk metal ($\alpha$-U). Similarly, for the energy of species such as Mo, Ru, Rh, Pd, and Tc, their elemental phases were considered Mo (bcc), Tc (hcp), Rh (fcc), Ru (hcp) and Pd (fcc). Additionally, for the energy of oxygen, we modeled oxygen molecules. Due to the well-known self-interaction error using DFT \cite{droghetti_predicting_2008}, an energy correction was applied using the suggested value from the fitted elemental-phase reference energies \cite{stevanovic_correcting_2012}.

The chemical potential $\mu_i$ includes the zero-point vibrational energy and the temperature and pressure dependence of the chemical potential. By neglecting the zero-point vibration and the pressure effects, the term $\mu_i$ can be ignored. This approximation is reasonable, particularly for defect elements at zero temperature \cite{Jean_2001_PRB,Zhu_2010_PRB,Reuter_2001_PRB,Minki_2012,zhou2025impacts}. However,it has to be noted that the oxygen chemical potential is important when considering the hyperstoichiometric UO$_2$ \cite{COOPER_2018}. Thus, for defect elements, the third term in Equation~(\ref{eq1}) is just the DFT-calculated energy for each species. Furthermore, at equilibrium, the chemical potentials of U and O are related by the formation enthalpy of bulk UO$_2$:
\begin{equation}\label{eq2}
    \mu_U + 2\mu_{O} = \Delta H_{UO_2}
\end{equation}  
where $\Delta H_{UO_2}$ is the formation enthalpy per formula unit as obtained from the DFT calculation. $\Delta H_{UO_2}$ is always negative to avoid decompose into $\alpha$-U and O molecule. The values of $\mu_U$ and $\mu_{O}$ depend on the stochiometry of UO$_2$: in the U-rich condition, $\mu_{U}=0$ and $\mu_{O}^{min} = \frac{1}{2} \Delta H_{UO_2}$, and in the O-rich condition, $\mu_{O}=0$ and $\mu_{U}^{min}= \Delta H_{UO_2}$). Throughout this work, we use the O-rich condition. Finally, to understand the charge transfer between the defects, the isosurfaces of charge density were visualized using the VESTA package \cite{Momma_vesta}.

%%%%Results
\section{Results and Discussion}
 
\begin{figure}[h!]
    \includegraphics[width=0.9\columnwidth]{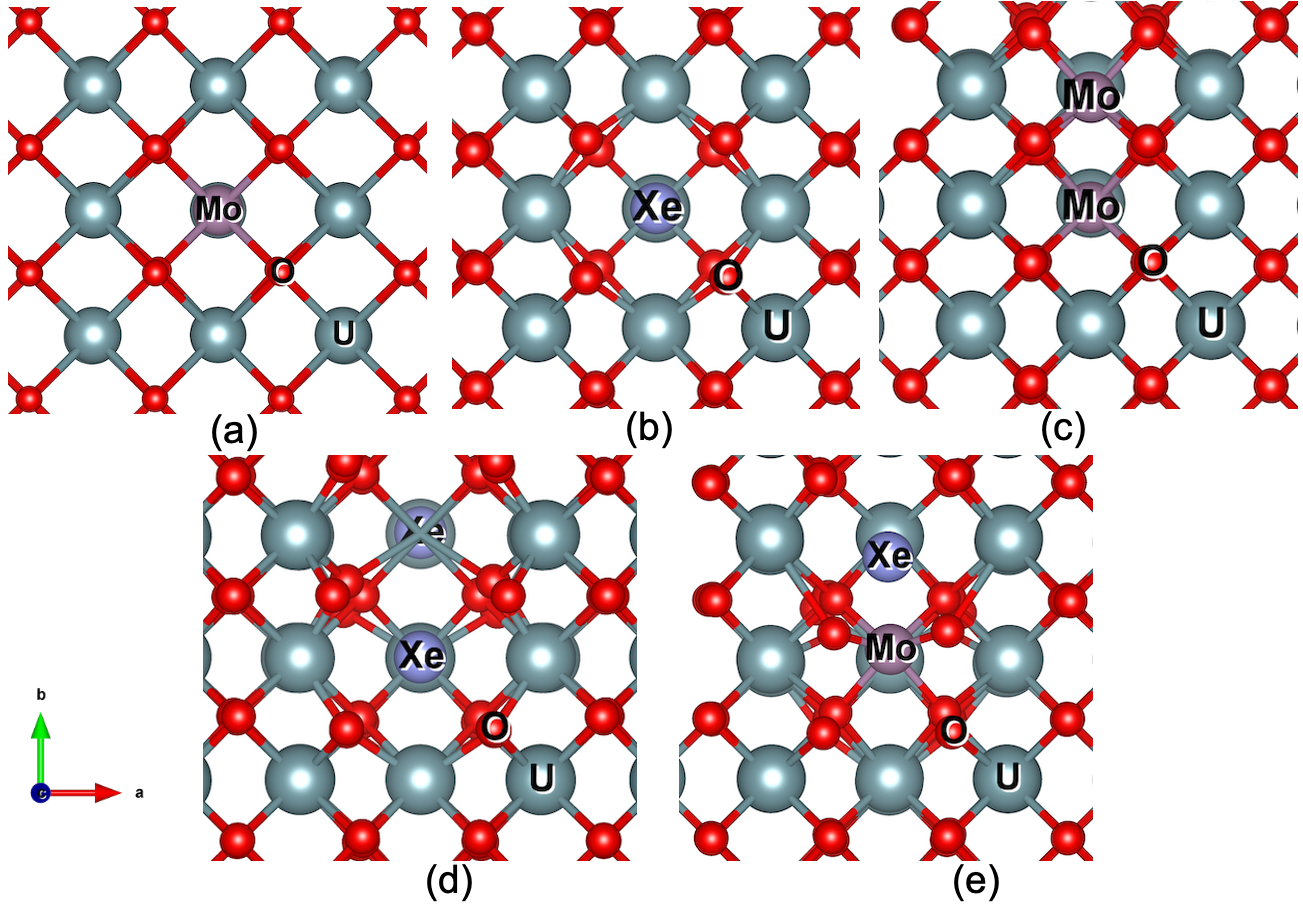}
    \caption{Relaxed structure of the Xe-Mo pair in UO$_2$ fuel with atoms having been substituted in the U site (a) Mo, (b) Xe, (c) two Mo atoms, (d) two Xe atoms, and (e) one Xe atom combined with one Mo atom.}
     \centering
    \label{Fig3.png}
\end{figure}

In this study, we examined five different types of defect pairs in UO$_2$, denoted as Xe-M (where M represents Mo, Ru, Rh, Tc, and Pd). To calculate the formation energy, we conducted structural optimization on UO$_2$ by substituting the U sites
with one metal atom, one Xe atom, two metal atoms, two Xe atoms, and one metal atom combined with one Xe atom. We limited our investigation to the substitution of defects in the U vacancy, as it is the most energetically favorable trap site for several fission products \cite{Nerikar_2009, BRILLANT_2011, Minki_2012}, due to the significant mismatch in atomic sizes between O and Xe, and between O and M. The relaxed structure of the Xe-Mo defect pair, used to calculate its formation energy, is presented in Fig.~\ref{Fig3.png}. Upon analyzing the relaxed structure, we observe that apart from the distorted nearest O atoms, when two Xe atoms are substituted in nearby U sites, one of them moves away from its original position, as shown in Fig~\ref{Fig3.png}(d), (a U atom was removed from Figure~\ref{Fig3.png}(d) to show the moved Xe atom). For the sake of brevity in this article, the relaxed structures of all other Xe-M pairs are provided in the supplementary information (SI-Fig 1).

To gain a deeper understanding of the observed defect pair structure, we conducted a comprehensive analysis of the charge density profiles for all Xe-M pairs. The charge density difference for the Xe-Mo pair is illustrated in Fig.~\ref{Fig3.png}, and the remaining cases are provided in the SI-Fig 2. In the isosurface regions, yellow denotes charge accumulation, while blue signifies charge depletion. The isosurface level was kept constant at a value of $4.64\times10^{-2}$~$e $Bohr$^{-3}$ for the overall charge density analysis. Our findings reveal significant charge depletion around Mo dopant sites, a trend consistent across other dopants (Xe and various metals). Additionally, when a single metal atom substitutes at a uranium site, there is a symmetrical accumulation of charge on the nearest oxygen atoms for most metals, except for Rh and Pd. This exception can be attributed to the highly filled valence orbitals in Rh and Pd. However, in the presence of Xe, charge accumulation extends to the next-nearest neighbors, likely due to the dangling bonds at these locations. The most significant charge redistribution is observed in cases where both Xe and metal atoms are present. These findings necessitate further investigation of the Bader charges on the dopant atoms and the host lattice.  

\begin{figure}[h!]
    \centering
    \includegraphics[width=0.9\columnwidth]{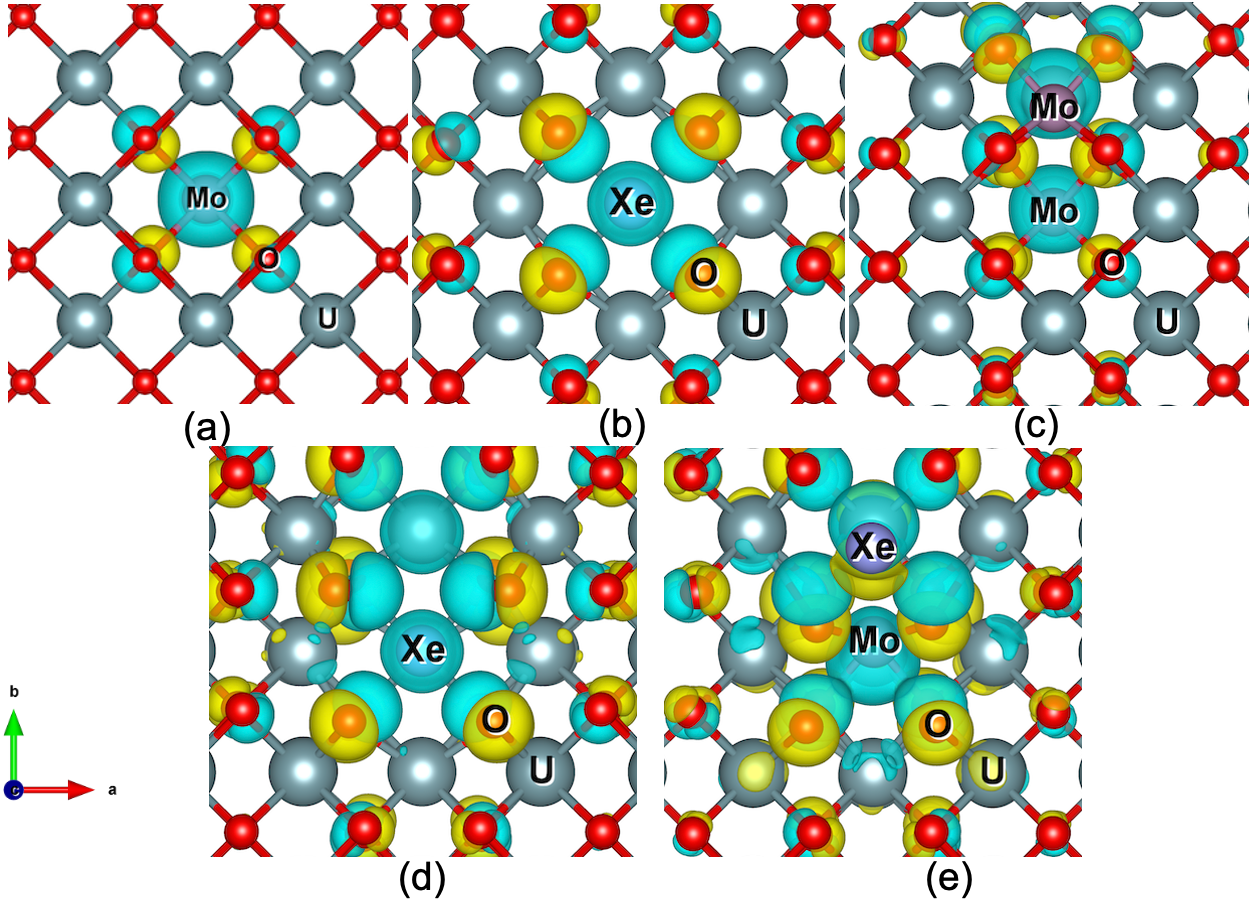}
    \caption{Charge density difference plot of the Xe-Mo pair in UO$_2$ fuel with atoms having been substituted in the U site (a) Mo, (b) Xe, (c) two Mo atoms, (d) two Xe atoms, and (e) one Xe combined with one Mo atom.}
    \label{Fig4.png}
    
\end{figure}

The Bader charges on various atoms for all possible defect configurations are listed in SI Table 1, where (q) denotes the Bader charge present on each atom. We observe a consistent pattern in the Bader charges for U and O, with U exhibiting values around 11.31 to 11.41 and O around 7.26 to 7.30, across different defect configurations. However, an intriguing trend is observed when examining the Bader charges for Xe and various metal atoms. The analysis reveals that the Bader charge on the Xe atom is relatively higher, indicating that Xe, as a noble gas, retains a significant amount of its electronic charge. Whereas, metal atoms show varying degrees of electron transfer, with Mo showing the most significant charge reduction and Pd retaining the highest charge. Specifically in the case of M-Xe-UO$_2$ configurations, the Mo-Xe-UO$_2$ system exhibits the lowest metal charge, signifying a strong electron transfer from Mo. This significant charge depletion in Mo results in notable charge delocalization, which is corroborated by the charge difference density results, as illustrated in Fig ~\ref{Fig3.png}. This charge delocalization likely plays a critical role in stabilizing the structure. These finding provide a crucial insights into the electronic interactions and charge distributions in M-Xe pair formation, highlighting the unique behaviour of Mo.

%\subsection{Formation Energy of the Xe-Metal pair}
To understand the formation energy of Xe-metal pairs, we first calculated the formation energy of single defects. The single defects in UO$_2$ were constructed by replacing one U atom with either a Xe, Mo, Ru, Rh, Pd, or Tc atom. The relaxed-structure schematics for the single-atom Xe and Mo substitutions are shown in Fig.~\ref{Fig3.png} (a), with all other single-defect relaxed structures being provided in the SI. The formation energy in an O-rich environment, as calculated using Equation~\ref{eq1} for all single defects considered in this work, is shown in Fig.~\ref{Fig4.png}. Among all the single defects analyzed, the Mo atom exhibits the lowest formation energy, indicating that the smallest energy penalty is incurred when substituting Mo at a U site. This is followed by Tc, Ru, Rh, Pd, and Xe. It is notable that the formation energy for single Mo and Tc atoms in UO$_2$ is negative, indicating a solution in the lattice. However, given the experimental evidence of NMP formation, it is reasonable to deduce that the presence of other metals makes the formation of a separate phase thermodynamically more favorable. This suggests a complex interaction that warrants further computational investigation using larger length scale modeling. Following the analysis of single defects, we calculated the formation energies of dimer defects of the same species, with two atoms of Xe, Mo, Ru, Rh, Pd, or Tc being substituted at two neighboring U sites. The calculated formation energies of the dimers are listed in Fig.~\ref{Fig4.png} (b) (denoted as ``coupled''), in comparison with the formation energies measured when the two defects are spaced infinitely far apart (``isolated''). The total formation energy of the two isolated defects was estimated by summing together the formation energies of each individual defect.

Figure~\ref{Fig4.png} (b) clearly demonstrates that coupled dimers are energetically more favorable than dispersed atoms. Specifically, the Xe$_U$-Xe$_U$, Ru$_U$-Ru$_U$, Pd$_U$-Pd$_U$, and Ta$_U$-Ta$_U$ pairs show a reduction in formation energy when in close proximity. In contrast, the changes in formation energy for Rh$_U$-Rh$_U$ and Mo$_U$-Mo$_U$ are marginal. These results indicate that metal atoms tend to form metallic clusters in UO$_2$ fuel, rather than existing as dispersed atoms. Finally, we computed the formation energies of defect pairs involving Xe and various metals (Mo, Ru, Rh, Tc, and Pd) and compared these with the formation energies pertaining to when the metal and Xe atoms are isolated, as illustrated in Fig.~\ref{Fig4.png} (c). Our results demonstrate that the formation energies of Xe-metal pairs are consistently lower than those of isolated defects, suggesting Xe-metal pair defects to be energetically favorable. Among the metals studied, the Xe-Mo pair exhibited the lowest formation energy, followed sequentially by Tc, Ru, Pd, and Rh. 
Furthermore, we calculated the binding energy of Xe-metal pairs, as illustrated in Fig.~\ref{Fig4.png} (d). The results suggest that Xe-metal interactions are energetically favorable, with the highest binding energy observed for the Xe-Pd pair, followed by the Xe-Mo pair. Charge transfer analysis indicated that the charge transfer between Pd and Xe is negligible. Consequently, we attribute the substantial binding energy in the Xe-Pd system primarily to minimal mismatch between the strain field, as Pd as the closest atomic radii as that of Xe, among the five metal precipitates studied. This could potentially reduce the overall energy of the system, leading to a higher binding energy. In the case of the Xe-Mo pair, the binding energy arises from a combination of local charge redistribution and the change of strain field. In summary, the strong binding energies for the Xe-metal pairs can be attributed to a synergistic effect of charge transfer and change of strain field upon the formation of the Xe-metal pairs. This finding indicates that, of the metals investigated in this study, Pd and Mo are the most likely candidates to form stable pair structures with Xe.

\begin{figure}[h!]
        \centering
        \includegraphics[width=0.9\columnwidth]{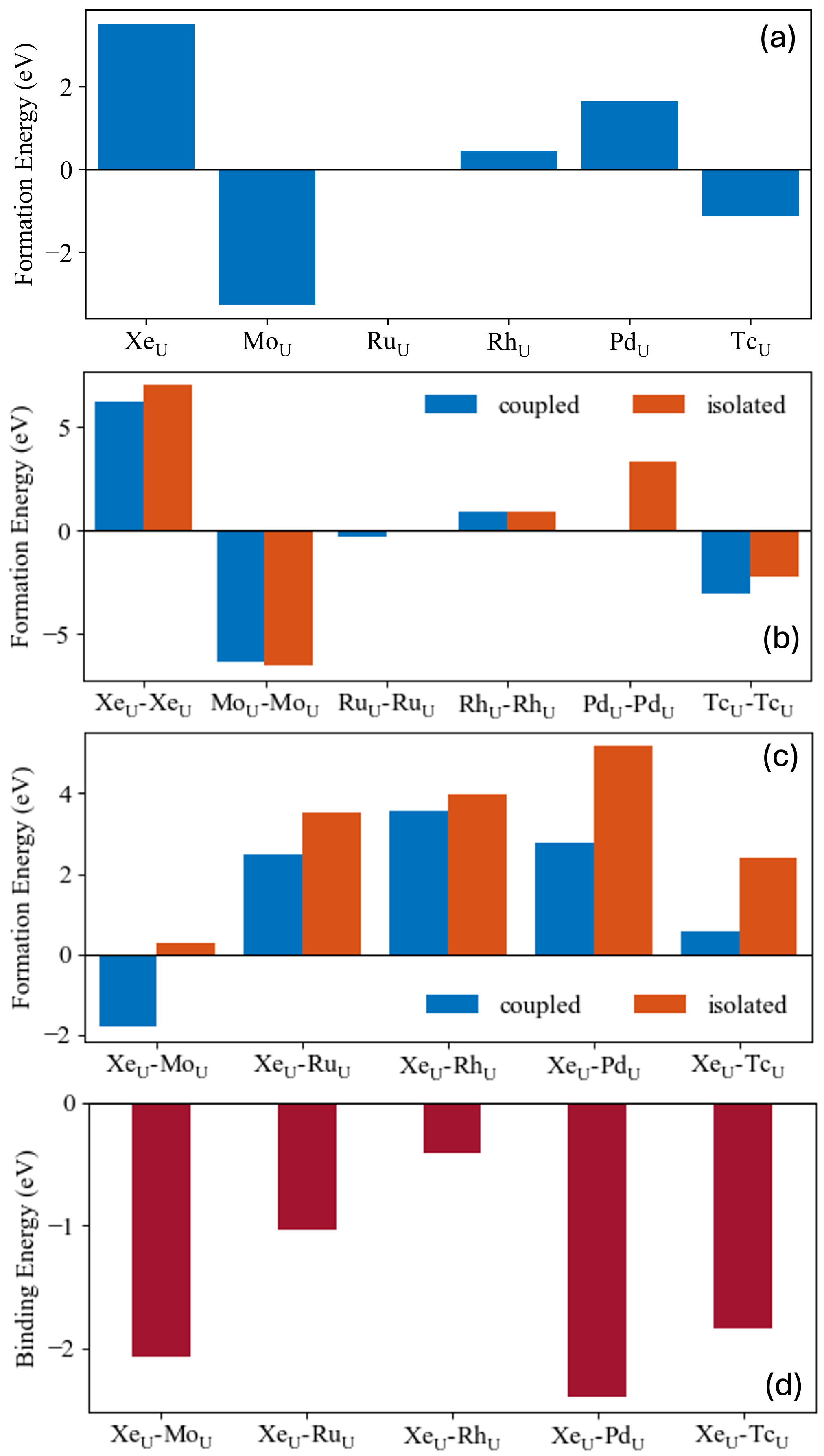}
        \caption{(a) Calculated formation energies of single defects at the U site, in the following order: Xe, Mo, Ru, Rh, Pd, and Tc. (b) Pair formation energies of dimers of the same species (blue), as compared to when these defects are isolated (orange). (c) Pair formation energies of dimers of the Xe-metal (blue), as compared to when these defects are isolated (orange).(d) Binding energy calculated by taking the energy difference for separated and coupled Xe-metal pairs}
       \label{Fig5.png}
    \end{figure}

Note that while the Xe-M pair formation energies can serve as an important milestone for studying the formation of clusters and precipitates, a multiscale modeling approach will be necessary to investigate the stability and morphology of the pair cluster formation as observed in Fig.~\ref{Fig1.png}(b). Simulation of the Xe-M clusters requires a robust description of the interatomic energies (e.g., binding or formation energies of clusters as a function of size and solute type) with \emph{ab initio} accuracy for a wide variety of cluster configurations. Specifically, a combination of DFT, mean-field, and coarse-grained approaches that involve cluster dynamics and statistical sampling \cite{CD_review} will be important for capturing Xe-M cluster evolutions. These evolutions can be captured by determining the rate processes of multiple cluster reactions (e.g., absorption, desorption, and re-solution) and diffusion events, and these rate processes will be used to parameterize multiscale models. The cluster dynamics model and statistical sampling have seen successful applications for investigating the stability and evolution of precipitates \cite{LEPINOUX20091086} and solute/defect clusters \cite{CD_MNS,JIN201833,Cr_cluster} in structural alloys. The integrated multiscale approach is currently beyond the scope of this study; future research will aim to incorporate this approach for a more comprehensive understanding.

\section{Conclusion}
In conclusion, we employed DFT+$U$ calculations to unravel the fundamental mechanism underlying the formation of Xe-M pair structures in spent UO$_2$ fuel samples from BR3. Our study calculated the formation energies and binding energies of five Xe-metal pairs, both in close proximity and in isolation. The results revealed that the formation energies of Xe-metal pairs are consistently lower than those of single defects, indicating enhanced stability for paired configurations. The binding energy of the Xe-M pairs follows the order of Xe-Pd $>$ Xe-Mo $>$ Xe-Tc $>$ Xe-Ru $>$ Xe-Rh, clearly highlighting that Pd and Mo are the most favorable metal for forming stable structures with Xe among the five metals considered. We attribute the stability of the metal-Xe pair to a synergistic effect of charge transfer and the change of strain field upon the formation of the Xe-metal pairs. Also, the APT analysis suggests that the concentration of Mo, Pd and Ru is relatively higher in the NMPs, indicating that these elements likely govern the fundamental mechanisms. This elucidation of the pair formation mechanism significantly advances our understanding of fission product behavior in nuclear fuels. Future energy-dispersive X-ray spectroscopy analyses of the BR3 samples are planned to further substantiate these findings through experimental validation. These insights enhance our comprehension of defect interactions in nuclear fuel matrices.

\section{ACKNOWLEDGMENTS}
This work was supported through Idaho National Laboratory (INL)'s Laboratory Directed Research and Development (LDRD) Program under U.S. Department of Energy (DOE)-Idaho Operations Office Contract DE-AC07-05ID14517. The authors also acknowledge that this research made use of the resources of the High Performance Computing Center at INL, which is supported by the DOE Office of Nuclear Energy and the Nuclear Science User Facilities under contract no. DE-AC07-05ID14517. The U.S. Government retains and the publisher, by accepting the article for publication, acknowledges that the U.S. Government retains a nonexclusive, paid-up, irrevocable, world-wide license to publish or reproduce the published form of this manuscript, or allow others to do so, for U.S. Government purposes.
 \bibliography{UO2}

% Produces the bibliography via BibTeX.

\end{document}